\documentclass[sigconf]{acmart}

\usepackage{graphicx}
\usepackage{xcolor}
\usepackage{xspace}
\usepackage{comment}
\usepackage{framed}
\usepackage{listings}
\usepackage[strict]{changepage}
\usepackage{color}
\usepackage{listings}
\usepackage{wrapfig}

\definecolor{dkgreen}{rgb}{0,0.6,0}
\definecolor{gray}{rgb}{0.5,0.5,0.5}
\definecolor{mauve}{rgb}{0.58,0,0.82}

\AtBeginDocument{%
  }

\setcopyright{acmcopyright}
\copyrightyear{2022}
\acmYear{2022}
\acmDOI{XXXXXXX.XXXXXXX}

\acmConference[ESEM '22]{Make sure to enter the correct
  conference title from your rights confirmation emai}{September 19--23,
  2022}{Helsinki, Finland}
\acmPrice{15.00}
\acmISBN{978-1-4503-XXXX-X/18/06}

\newcommand{\cddguided}{CDD-guided refactorings\xspace}

\newcommand{\totalSurveyParticipants}{133\xspace}
\newcommand{\totalSurveyZup}{73\xspace}
\newcommand{\totalSurveyOnline}{60\xspace}

\newcommand{\RQone}{Do CDD-guided refactorings improves code readability, according to professional software developers?\xspace}
\newcommand{\RQtwo}{Do CDD-guided refactorings improves code readability, according to the state-of-the-art readability model?\xspace}

\begin{document}

\title{To What Extent Cognitive-Driven Development \\ Improves Code Readability?}

\author{Leonardo Barbosa}
\email{leonardopfb@gmail.com}
\orcid{XXXX-XXXX-XXXX}
\affiliation{%
  \institution{UFPA}
  \city{Belém}
  \state{PA}
  \country{Brazil}
}

\author{Victor Hugo Santiago}
\email{victor.santiago@ufpa.br}
\orcid{XXXX-XXXX-XXXX}
\affiliation{%
  \institution{UFPA}
  \city{Belém}
  \state{PA}
  \country{Brazil}
}

\author{Alberto Luiz Oliveira Tavares de Souza}
\email{alberto.tavares@zup.com.br}
\orcid{XXXX-XXXX-XXXX}
\affiliation{%
  \institution{Zup Innovation}
  \city{Belém}
  \state{PA}
  \country{Brazil}
}

\author{Gustavo Pinto}
\email{gustavo.pinto@zup.com.br}
\orcid{XXXX-XXXX-XXXX}
\affiliation{%
  \institution{UFPA \& Zup Innovation}
  \city{Belém}
  \state{PA}
  \country{Brazil}
}

\renewcommand{\shortauthors}{Barbosa et al.}
\renewcommand{\shorttitle}{To What Extent Cognitive-Driven Development Improves Code Readability?}

\begin{abstract}
Cognitive-Driven Development (CDD) is a coding design technique that aims to reduce the cognitive effort that developers place in understanding a given code unit (e.g., a class). By following CDD design practices, it is expected that the coding units to be smaller, and, thus, easier to maintain and evolve. However, it is so far unknown whether these smaller code units coded using CDD standards are, indeed, easier to understand. In this work we aim to assess to what extent CDD improves code readability. To achieve this goal, we conducted a two-phase study. We start by inviting professional software developers to vote (and justify their rationale) on the most readable pair of code snippets (from a set of 10 pairs); one of the pairs was coded using CDD practices. We received \totalSurveyParticipants answers. In the second phase, we applied the state-of-the art readability model on the 10-pairs of \cddguided. We observed some conflicting results. On the one hand, developers perceived that seven (out of 10) \cddguided were more readable than their counterparts; for two other \cddguided, developers were undecided, while only in one of the \cddguided, developers preferred the original code snippet. On the other hand, we noticed that only one \cddguided have better performance readability, assessed by state-of-the-art readability models. Our results provide initial evidence that CDD could be an interesting approach for software design.
\end{abstract}




\settopmatter{printfolios=true}

\maketitle

\section{Introduction}

Cognitive-Driven Development (or CDD for short) is a novel coding design technique that aims to reduce code complexity by limiting the number of language constructs that could be used at once in a given source code file~\cite{CDD:ICSME:2020,CDD:ENASE:2021,CDD:ICEIS:2022}. CDD aims at developing strategies that reduce the developer cognitive load. %These strategies are not fixed and can be tailored for every development team. CDD recognizes the human limitation to guide software development, prioritizing the understanding and quality metrics.

Instead of being based on anecdotal experience, CDD has its roots in two well-known psychology theories: the Magical Number Seven Theory~\cite{miller1956magical} and the Cognitive Load Theory (CLT)~\cite{sweller1988,sweller2010}. 
In Miller's work~\cite{miller1956magical} known as ``Magical Number Seven'', it is explained that we probably have a hard limitation in simultaneous processing information. Experimental studies.\cite{miller1956magical} have suggested that humans generally hold only seven plus or minus two information units in short-term memory. 
CLT~\cite{sweller1988,sweller2010}, on the other hand, explains that any material has an intrinsic complexity depending on the amount and arrangement of the elements that compose it. CLT is an instructional design theory that reflects our ``cognitive architecture'', an important aspect to presenting information at a pace that learners can fully understand. According to Sweller~\cite{sweller1988}, knowing the number of information elements and their interactivity is crucial to support learners.

In terms of source code, it may not be a surprise that a high number of information elements, e.g., control structures and language constructs, can harden ones understanding. CDD focuses on directing developers to create and maintain code units respecting the limited human cognition capacity. Therefore, the more elements are concentrated in the same code unit, the greater the effort developers will have to place to understand it. While the practice of CDD could indeed ease software maintenance (since CDD-guided source code uses less code elements), little is known whether the code that was developed using CDD has also better readability.
%Although there is some early work on CDD (see more at Section~\ref{sec:cdd-related}), none of them focus on understanding whether the coding changes that are made guided by CDD actually improves code readability. 
Therefore, the goal of this work is to provide answers to the following research questions:

\begin{itemize}
    \item[\textbf{RQ1.}] \RQone
    \item[\textbf{RQ2.}] \RQtwo
\end{itemize}

To answer these questions, we performed a two-phase study. To answer \textbf{RQ1}, in the first phase of this work we started by surveying  \totalSurveyParticipants professional developers. We present to them 10 pairs of code snippets (pre and pos \cddguided) and ask our subjects to vote on which code snippet they believe is more readable. We also asked them to offer their rationale behind their vote. To answer \textbf{RQ2}, in the second phase, we leverage the state-of-the-art readability model proposed by Posnett et al.~\cite{Posnett:MSR:2011}. For each pair of code, we studied whether the \cddguided might have, indeed, improved code readability. 
Among the 10 pairs of code snippets evaluated, we observed that in seven of them, our participants concur that the \cddguided improved readability (in two pairs the results were balanced, while in just one the participants preferred the original code). However, when performing the state-of-the-art readability model, we noticed a different figure. In this case, just one of the \cddguided was considered as more readable, according to the model. This work makes the following contributions.

\begin{enumerate}
    \item We curated a set of 10 \cddguided (along with their original versions), created by professional software developers;
    \item We conducted a survey with \totalSurveyParticipants professional software developers to assess their perception about the 10 pairs of code snippets.
    \item We applied the state-of-the-art readability model in the 10 pairs of code snippets. 
    \item We made available all data used in this research, to facilitate further replication.
\end{enumerate}

\section{How does CDD work?}

Complexity is part of the software and CDD recognizes that the intrinsic load effects from code units are different for people. Suppose we have two developers analyzing the same program. They will likely disagree on the difficulty of understanding. Nonetheless, when code needs to evolve, the difference in observed complexity will likely impact the maintenance cost. CDD aims to create an unified concept for intrinsic complexity for the code units.

The first step of the CDD approach is to define the ``Intrinsic Complexity Points'' (ICPs)~\cite{CDD:ICSME:2020}, which are the elements inside the code that can affect the understanding according to their usage frequency. Example of such elements are for instance: code branches (\texttt{if-else}, \texttt{loops}, \texttt{when}, \texttt{switch-case}, \texttt{do-while}, \texttt{try-catch} and etc.), functions as an argument, contextual coupling (i.e., coupling with specific project classes), and inheritance. CDD is not limited to those code elements, though. Indeed, developers can include any other code elements, from SQL instructions, to annotations, to assertions, etc. Any programming construct that the team consider relevant can be considered as ICP. 

After the selection of ICPs, a constraint should be defined for all code units. CDD can contribute to a practical perspective of what makes understanding compromised for a given context. For example, how many \texttt{if} statements should I have in a given code unit? 
As everyone is involved in creating the code, a complexity limit can be defined collaboratively for the code units regardless of the specialization degree of developers. Therefore, development teams can adopt the CDD method according to their interest, the project nature and experience, i.e., the cognitive complexity limit needs to be customized; this is also true for the elements included in this constraint. The main reward of this strategy is that the code units can be kept under a limit even with the exponential growth of the software complexity.

%A simple constraint known by team members can take advantage of the fragility of the definition of what is or is not complex. Therefore, the same notion of complexity can influence the developers during programming.

%\gnote{descrever que CDD é flexivel, que times podem usar de acordo com seu interesse e experiencia.}

\section{Research Methodology}

This work was built upon the work of Pinto et al.\cite{CDD:ENASE:2021} and Santos and Gerosa~\cite{Santos:ICPC:2018}. Regarding the work of Pinto et al.\cite{CDD:ENASE:2021}, we relied on the refactoring produced by developers from industry that took part of the experiment (more in Section~\ref{sec:refactoring}). After cleaning and filtering representative refactoring examples, we adopted the survey approach presented by Santos and Gerosa~\cite{Santos:ICPC:2018} to gather developers' opinion on which code snippets have better readability  (more in Section~\ref{sec:survey}). We though extend the Santos and Gerosa work by applying the state-of-the-art readability model, proposed by Posnett et al.~\cite{Posnett:MSR:2011}  (more in Section~\ref{sec:model}).

\subsection{Curating refactoring examples}\label{sec:refactoring}

%\gnote{esta faltando algo aqui? o que uma pessoa com acesso aos dados do Victor deveria fazer pra ter os mesmo 10 pares de exemplo que temos? está *tudo* contemplado aqui?}

Pinto et al.~\cite{CDD:ENASE:2021} carried out an experimental study to determine whether refactoring using cognitive driven constraints leads to better software than traditional style refactoring. The authors have conducted the experiment in an industrial setup and evaluated software quality via software quality metrics. Nonetheless, since the industry participants performed several scattered changes, the authors were unable to understand the impact of isolated CDD-guided refactorings on code readability. We started by selecting refactorings that were produced from this research. The authors invited 18 industry participants to refactor three classes of two well-known Java projects, namely SSP\footnote{https://github.com/Jasig/SSP} and feign\footnote{https://github.com/OpenFeign/feign}. The participants were divided in two groups: the first group performed their refactorings using their own intuition, whereas the second group followed a disciplined CDD approach. To carry out their work, both groups should meet the following requirements: 

\begin{enumerate}
    \item The original project's packages should not be modified;
    \item New packages should not be created during refactorings;
    \item Automated tests should continue working without changes, and;
    \item Public, protected, or package private methods from original class should not be modified. 
\end{enumerate}

Additionally, the participants in the CDD group had to adhere with the following set of ICPs: code branches (i.e., \texttt{if-else}, \texttt{loops}, \texttt{when}, \texttt{switch/case}, \texttt{do-while} blocks), exception handling (i.e., \texttt{try-catch} blocks), functions as an argument, and contextual coupling. It is important to highlight that the participants of the CDD group were free to define a feasible complexity constraint for the code units based on possible ICPs. According to the authors, a single recommendation to define this constraint is that it could be equivalent to twice the number of ICPs chosen for accounting.

This group was encouraged to follow a progressive reduction strategy of ICPs. For instance, such participants were trained to manually identify the aforementioned set of ICPs in the classes to be restructured. After that, the participants had to refactor the classes in order to reduce the total number of ICPs to satisfy the constraint previously defined. According to the authors, their hypothesis was that the CDD-guided refactorings could lead to more quality in terms of static metrics. The authors though used CK metrics~\cite{CK:TSE} (namely CBO, LCOM, RFC and WMC) as a proxy of readability.

%through know refactoring techniques (composing methods, moving features between objects, simplifying conditional expressions and so on).

%Os trechos são de três classes JournalEntryServiceImpl e EarlyAlertServiceImpl do projeto SSP https://github.com/Jasig/SSP, e a classe RequestTemplate do feign https://github.com/OpenFeign/feign.

In this work we extend the work of Pinto et al.~\cite{CDD:ENASE:2021} by focusing on understanding to readability of isolated CDD-guided refactorings. We asked the authors their dataset, which they gently provided. The tool named Meld~\footnote{https://meldmerge.org/} was adopted to compare files and visualize the changes performed by the participants.

When we started visualizing the code changes, we quickly noticed that not all code changes were performed following CDD guidelines. For example, some changes were targeted to streamline methods and remove code duplication. We then focused our search on ``isolated CDD-guided refactorings'', that is, code changes that we could assure that were guided by CDD. We used the following approach to identify the ``isolated CDD-guided refactorings'': 

\begin{enumerate}
    \item We searched for code changes that could match with the CDD requirements employed. For example, if the changes tried to reduce the \texttt{if}, \texttt{for} or \texttt{try-catch} statements, chances are that they were guided by CDD;
    \item We sought to observe indications that an intrinsic complexity constraint was defined to guide the refactorings. For example, when the target classes for refactorings were changed and the new classes were created following a clear definition for a complexity limit considering the basic control structures aforementioned;
    \item We focused on code changes that adhere with the seminal refactoring definition: code changes that preserve external behavior~\cite{opdyke1990}. If the refactored source code introduced additional code elements that could potentially change the program's behavior, we discarded that code change.
\end{enumerate}

This process was independently performed by two researchers and was revised by a third researcher. This process took about six weeks. We limited the number of CDD-guided refactorings to 10 pairs, 20 at total (10 before and 10 after the refactoring). This number is similar to the number of examples used in the work of Santos and Gerosa~\cite{Santos:ICPC:2018}, which chose 11 examples.

The selected pairs have are fairly small (on average, 25 lines of code); more at Table~\ref{tab:model}). This is aligned with the goals of our work. For our \textbf{RQ1}, we intended to ask developers about their perceptions of each refactoring example --- and having a high number of examples would potentially tire the respondent (which in turn could lead to survey dropouts). For \textbf{RQ2}, our applied model have better performance for small code snippets (the authors conducted their work with code snippets ranging from 4 to 11 lines of code; code snippets with 200+ lines of code were all ranked as less readable.

\subsection{Designing and Deploying the survey}\label{sec:survey}

Our survey design was inspired by the work of Santos and Gerosa~\cite{Santos:ICPC:2018}. In their work, the authors asked developers about their perceptions regarding 10 pairs of code snippets. Since the authors were interested in visually assessing the pairs of code examples, they were not not able to rely on tools such as Google Form or Survey Monkey, because they do not offer such visual features. Instead, they opted to built their own survey tool. The survey participants were mostly composed by students (55 respondents), in addition to seven professional developers. In the next, we discuss how our work differs from theirs. Also, our survey was based on the recommendations of Kitchenham et al.\cite{kitchenham2008personal}, we followed the phases prescribed: planning, creating the questionnaire, defining the target audience, evaluating, conducting the survey, and analyzing the results. We discuss each one of these phases next.

\vspace{0.2cm}
\noindent
\textbf{Planing.} We identified 12 representative refactoring examples. However, as aforementioned, we had to discard two examples to avoid occupying too much the participants. Therefore, we group the refactoring examples into categories and discarded those from most common categories. We ended up with 10 pairs of code snippets.

Moreover, we decided not to mention that the refactorings were CDD-guided since this could lead to biases for (or against) CDD. Instead, we just asked the participant about their preference without further information about the pairs of examples. We also decided to randomly alternate the sides of the examples, otherwise participants could naively favor one of the sides (left or right). Finally, since readability is a subjective concept and that different factors can influence code readability, before presenting to developers the pairs of examples, we present to them our definition of readability: ``Readability is a human judgment about the cognitive effort required to understand a source code. Visual features such as spacing, indentation, capitalization, function names, language constructs, among others, can affect the readability and, thus, the understanding of a code snippet.''. This definition is inspired by the works of Oliveira et al~\cite{Oliveira:ICSME:2020}, which describes that readability is ``what makes a program easier or harder to read and apprehend by developers''.

\begin{figure}[ht]
    \centering 
    \includegraphics[width=\linewidth, clip=true, trim= 1px 1px 1px 1px]{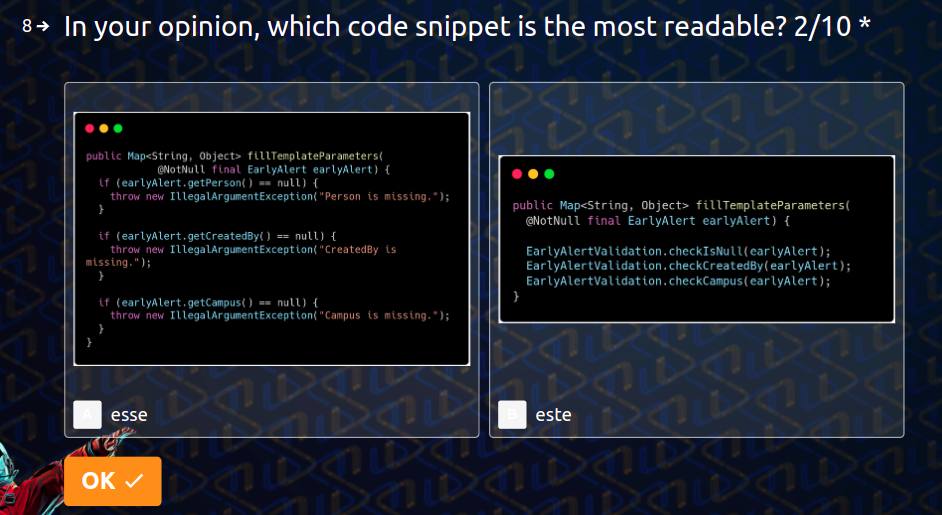}
    \caption{Most readable snippet selection screen.}
    \label{fig:survey}
\end{figure}

\vspace{0.2cm}
\noindent
\textbf{Creating the questionnaire.} 
In this work, we relied on TypeForm, which is an online survey service that allows dynamic forms. Using TypeForm, we were able to present developers with two images and ask them their preference. We used this feature to present our pairs of examples, as Figure~\ref{fig:survey} shows. Knowing that syntax highlight could improve code comprehension~\cite{hannebauer2018does}, different than Santos and Gerosa~\cite{Santos:ICPC:2018}, we used the \texttt{carbon.now.sh} tool to create code snippets with syntax highlight in dark mode (which offer better contrasts to the figures). After the participant vote of their preferable code snippet, we asked their rationale for this decision.

Our survey had 27 questions (25 were required, 12 were open). We asked five demographic questions (Technical Profile \{Developer, QA, Manager, etc\}, Technical Level \{Novice, Intermediate, Senior, Principal, etc\}, Age \{ $<$20y, 21--30y, 31--40y, 41--50y, $>$50y \}, Years of experience \{ $<$1y, 1--5y, 6--10y, 10--15y, $>$15y \}, Experience with Java \{1--10\}). Since we had to separate the questions regarding the refactorings preference from the developers rationale for their choice, we ended up with 20 questions for the 10 refactorings. In the end of the survey, we had two final open questions: one whether the participant had any questions/comments regarding our study and the other if she is interested in participating in a follow up interview. The estimated time to complete the survey was 10--15 minutes.

\vspace{0.2cm}
\noindent
\textbf{Evaluating the survey.} Before deploying the actual survey, we conducted a pilot with professional software developers and researchers. These developers and researchers were close personal contact to the authors of this work. The goal of this pilot survey was to assess the clarity of the questions and the quality of the refactorings examples. The participants of this pilot survey were instructed that their feedback regarding the questions and the code snippets was more important than the answers themselves. After a period of one week, we received 8 answers to the survey and a few comments requesting clarifications. We applied the suggestions and removed the 8 answers from the database.

\vspace{0.2cm}
\noindent
\textbf{Conducting the survey.} After incorporating improvements suggested in the pilot survey, we administered the actual survey. To do this, we created two online questionnaires. The first questionnaire was deployed at a large Brazilian software producing company, while the other questionnaire was shared in two social platforms: Twitter and LinkedIn.

For the \emph{first questionnaire}, one of the author work as a researcher in the company and had access to the company communication's channels. However, instead of sending the questionnaire to all $\sim$3.5k company's employees (which is against the company privacy culture), we shared the link of the questionnaire in the company's general Google space (similar to a Slack channel). Periodic reminders were sent at that same Google space. 

The \emph{second questionnaire} that was shared in social networks had exactly the same questions and options of the first one, but a different URL address. We decided to have two questionnaires because we had to report the perception of the company's employees. 
Given the nature of the approach we used to invite participants, we were unable to track the number of participants that received the questionnaire. However, Typeform tracks the number of access to the questionnaire. For instance, (255+328) participants opened the questionnaire and (210+201) started filling the questionnaire.
After a period of two weeks, we received a total of \totalSurveyParticipants responses, \totalSurveyZup for the first questionnaire (34.8\% of completion rate), and \totalSurveyOnline for the second one (30\% of completion rate). For both surveys, participation was voluntary. 

\begin{table}[h!]
    \centering
    \small
    \caption{Participants roles \& programming experience.} 
    \label{tab:exp}
    \begin{tabular}{ll|ccccc}
    \toprule
    Role & Population & $<$1y & 1--5y & 6--10y & 10--15y & $>$15y \\
    \midrule
    Development    & 125  &22  & 32 & 34 &19 & 18\\
    Management     & 6    & 1  & 0  & 2  & 1 & 2  \\
    Testing \& QA  & 1    & 0  & 0  & 1  & 0 & 0 \\
    Infrastructure & 1    & 0  & 0  & 1  & 0 & 0  \\
    \midrule
                   & 133  & 23 & 32 & 38 & 20 & 20  \\
    \bottomrule
    \end{tabular}
\end{table}

\begin{figure}[h]
    \centering 
    \includegraphics[width=0.8\linewidth, trim= 30px 00px 00px 0px]{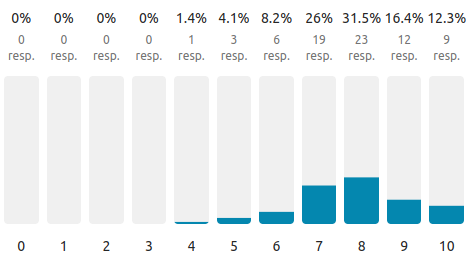}
    \caption{Java Experience.}
    \label{fig:java}
\end{figure}

\vspace{0.2cm}
\noindent
\textbf{Target audience.} The target audience of our work are professional software developers. Table~\ref{tab:exp} summarizes our participants demographics. %Among the respondents, 96\% work with software development, and the other 4\% are managers. 
Regarding their seniority, 9\% consider themselves as novice developers, 45\% are intermediate developers, and 33\% are senior developers. The remaining 13\% play different roles, such as Tech lead and C-level. As for their ages, 49\% have between 21 and 30 years old, 38\% have between 31 and 40 years old, 11\% have between 41 and 50 years old, and only 1.4\% have more than 50 years old. %\begin{wrapfigure}[8]{r}{0.6\columnwidth}
%    \centering
%    \vspace{-4ex}
%    \includegraphics[width=.8\linewidth, trim= 30px 00px 00px 0px]{figures/java-exp.png}
%    \vspace{-2ex}
%    \caption{\label{fig:java}Java experience.}    
%\end{wrapfigure}
As for their software development experience, 22\% have up to 1 year, 29\% have between 2 to 5 years of experience, 23.3\% have between 6 and 10 years, another 12\% have between 11 to 15 years, and 14\% have more than 15 years of software development experience. In a scale from 0 to 10, where 0 means no knowledge at all and 10 means being an expert, our participants averaged 7.8 in Java programming (Figure~\ref{fig:java}).

\vspace{0.2cm}
\noindent
\textbf{Analyzing the results.}
We performed both quantitative and qualitative analysis methods. For the quantitative method, we used the Chi-Square ($\chi^2$) statistical test. $\chi^2$ is a nonparametric test (i.e., it does not require normality). The test indicates whether our hypothesis holds or not. We defined the $\alpha$ level to the conventional level of 0.05. For the qualitative method, one of the authors followed qualitative coding techniques to categorize the respondents' perception. This approach was performed by one author and revised by another author. The most interesting observations are discussed throughout Section~\ref{sec:results} along with quotes from the survey. Among similar observations, we chose to quote only the one we considered the most representative for each case.

%, and our null and alternative hypothesis are then:
%https://www.jmp.com/en_be/statistics-knowledge-portal/chi-square-test/chi-square-goodness-of-fit-test.html
%\begin{itemize}
%    \item \textbf{Null hypothesis:} CDD-guided refactorings do not improve program readability, according to professional software developers.
%    \item \textbf{Alternative hypothesis:} CDD-guided refactorings do improve program readability, according to professional software developers.
%\end{itemize}
%\gnote{Complementar algo aqui?}

\subsection{Applying Posnett et al. readability model}\label{sec:model}

Although it is consensus that code readability is a subjective matter, in the last decades, researchers have proposed several metrics, tools, and models to assess code readability \cite{Posnett:MSR:2011,Buse:TSE:2009,CK:Scalabrino2017,MiQ:2018,Sivaprakasam:2012,Chung:2010}. 

In this work we leverage the model proposed by Posnett and colleagues~\cite{Posnett:MSR:2011}. This model is a simplified mix of two other approaches: the Buse model~\cite{Buse:TSE:2009} and the Hastead metrics~\cite{halstead1977elements}. Posnett and colleagues proposed a simplification of these two other works that consider only three variables: \emph{lines of code}, \emph{volume}, and \emph{entropy}. Volume and Entropy are two similar metrics. The author differs  \emph{volume} and \emph{entropy} as the follows: ``\emph{Entropy calculations depend on the relative distribution of the tokens/characters in the code body under consideration, with uniform distributions giving the highest entropy, and highly skewed distributions yielding lower entropy; whereas volume attempts to determine the number of bits needed to represent all operators and operands multiplied by the total number of tokens}.''

To calculate the Posnett~\cite{Posnett:MSR:2011} model, we used the Java-based implementation provided by Mendonça and colleagues~\cite{Mendonca:JSERD:2020}. To run the tool, we have to create the files for each code snippet (20 files at total). According to the tool documentation, ``\emph{The content of the file should be just a Java method body (and not an entire Java class)}''. If the source code contained a \texttt{class} definition, the tool raised an exception and finished execution. This limitation might be due to the fact that the seminal work of Posnett and colleagues~\cite{Posnett:MSR:2011} did not include class definition in their evaluated code snippets. Therefore, we could only calculate Posnett et al. readability model for methods (more details in Section~\ref{sec:rq2}).

\subsection{Dataset availability}

In order to foster replications of our work, we made available all data used in this research online at \url{https://bit.ly/3LG0eIm}. It includes 1) the curated sample of 10 pairs of code snippets, 2) the online questionnaire, and 3) the answers anonymized. We encourage others to replicate our work.

\section{Results}\label{sec:results}

We organize our results in terms of the research questions.

\subsection*{RQ1. \RQone}

%Without doing any statistics, we can see that the number of pieces for each flavor are not the same. Some flavors have fewer than the expected 200 pieces and some have more. But how different are the proportions of flavors? Are the number of pieces “close enough” for us to conclude that across many bags there are the same number of pieces for each flavor? Or are the number of pieces too different for us to draw this conclusion? Another way to phrase this is, do our data values give a “good enough” fit to the idea of equal numbers of pieces of candy for each flavor or not?

%To decide, we find the difference between what we have and what we expect. Then, to give flavors with fewer pieces than expected the same importance as flavors with more pieces than expected, we square the difference. Next, we divide the square by the expected count, and sum those values. This gives us our test statistic.

Table~\ref{tab:rq1} presents the overview of our survey results. As one could observe, for the majority of pairs, participants agreed that the \cddguided were more readable. In particular, P3, P10, P4, P2, and P6 were fairly well-voted (they acquired 93\%, 92\%, 89\%, 86\%, and 80\% of the votes, respectively). For two of the pairs (P1 and P7), we were unable to derive consensus. Finally, for only P5, participants voted against CDD. The results for and against CDD were statistically significant. Given this scenario, we were unable to reject null hypothesis.

\begin{table*}[t!]
    \centering
    \caption{Received Votes per Code Snippet Pair.} 
    \label{tab:rq1}
    \begin{tabular}{llccccc}
    \toprule
    Pairs & Short Description & Original & CDD-guided & \% Original & \% CDD-guided & $\chi^2$ \\
    \midrule
    P1 & Encapsulating error handling & 67 & 66 & 50\% & 50\% & 0.9309\\
    P2 & Encapsulating business rules & 19 & 114 & 14\% & 86\% & 0.0000 \\
    P3 & Concatenating logical expressions & 9 & 124 & 7\% & 93\% & 0.0000\\
    P4 & Extracting class & 14 & 119 & 11\% & 89\% & 0.0000 \\
    P5 & Listing all imports & 45 & 88 & 34\% & 66\% & 0.0001 \\
    P6 & Encapsulating business rules & 27 & 106 & 20\% & 80\%  & 0.0000\\
    P7 & Asserting conditions & 60 & 73 & 45\% & 55\% & 0.2596 \\
    P8 & Functional checking style & 95 & 38 & 71\% & 29\%  & 0.0000\\
    P9 & Extracting class & 44 & 89 & 33\% & 67\%  & 0.0000\\
    P10 & Encapsulating for loops & 10 & 123 & 8\% & 92\%  & 0.0000 \\
    \bottomrule
    \end{tabular}
\end{table*}

%\gnote{fazer uma descrição geral dos resultados, 2--3 paragrafos. em 7 de 10, CDD se da bem. em 2 casos, cdd empata (P1 e P7). e o P8 ele perde}

To better understand the reasons behind developers intuition of readability, after presenting each pair, we asked their rationale for their votes. We now present the results of the analysis of these comments. We summarize their opinions for each pair next. 

%\gnote{Victor e Leonardo: pensei somente agora que a gente deveria mencionar nos PNs de que foram a verão do CDD diminiu os ICPs}

\begin{figure}[h]
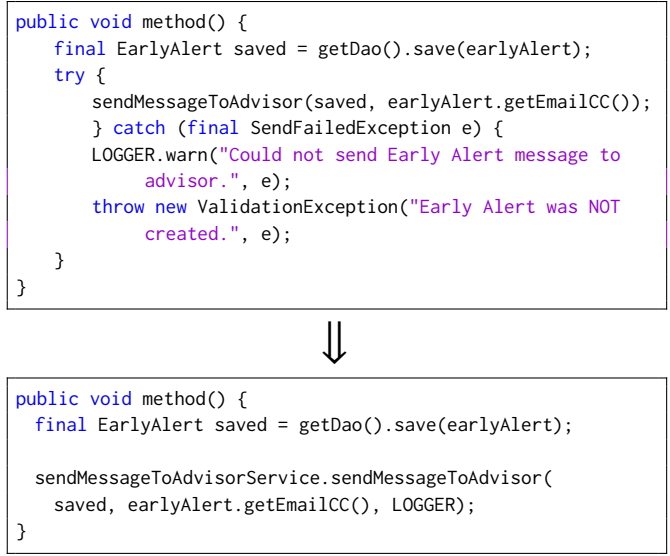

\begin{center}
\label{fig:p1}
\begin{minipage}{0.48\textwidth}
\begin{lstlisting}
public void method() {
    final EarlyAlert saved = getDao().save(earlyAlert);
    try {
        sendMessageToAdvisor(saved, earlyAlert.getEmailCC());
        } catch (final SendFailedException e) {
        LOGGER.warn("Could not send Early Alert message to advisor.", e);
        throw new ValidationException("Early Alert was NOT created.", e);
    }
}\end{lstlisting}
\end{minipage}

\scalebox{2}{$\Downarrow$}

\vspace{0.2cm}
\begin{minipage}{0.48\textwidth}
\begin{lstlisting}
public void method() {
  final EarlyAlert saved = getDao().save(earlyAlert);

  sendMessageToAdvisorService.sendMessageToAdvisor(
    saved, earlyAlert.getEmailCC(), LOGGER);
}
\end{lstlisting}
\end{minipage}

\vspace{-0.2cm}
\caption{P1: Encapsulating error handling}
\vspace{-0.2cm}
\end{center}
\end{figure}

\vspace{0.2cm}
\noindent
\textbf{P1: Encapsulating error handling (50\% / 50\%).}
In this example, the original code had a \texttt{try-catch} block handling the \texttt{SendFailedException} exception. The CDD-refactored version of this code extracted the exception handling responsibility and moved the \texttt{try-catch} block into a new method called \texttt{sendMessageToAdvisor}. Figure 3 shows this pair.
We noted a remarkable similarity in the votes; 66 participants preferred the original code, whereas other 66 preferred CDD-guided version. 
When analyzing the comments of this pair, we noticed that developers that favored the original code have a certain habit with this coding practice, as one respondent argued: ``\textit{Logging with \texttt{try-catch} is so common that I didn't even need to read the code to understand the purpose.}''. Another participant mentioned that ``\textit{there is more visibility of the actions performed in the code, what is being executed is more explicit.}"
%;"\textit{ It's the way I'm most used to doing it}"), demonstrating a certain habit with the model developed in the original code. 
On the other hand, the participants that favored the CDD-refactored version pointed out that the refactored code is closed to plain English, as one respondent mentioned: ``\textit{the code is shorter and with a language closer to the natural one}''. Other respondents mentioned that the code has a clear purpose: ``\textit{the removal of \texttt{try-catch} avoids distraction from the most important part of the code}", and ``\textit{Isolating error handling within a method gives me a partial view, according to the responsibility of each method}".
%;"\textit{The code is cleaner}";"") 
The decoupling leaves the most important part in the main code and therefore easier to understand, in addition to the method created having a name well suited to its purpose makes it easier to understand.

\begin{figure}[h]
\begin{center}
\begin{minipage}{0.48\textwidth}
\begin{lstlisting}
public EarlyAlert create(@NotNull final EarlyAlert earlyAlert) {
  if (earlyAlert.getPerson() == null) {
    throw new ValidationException("Person is missing.");
  }

  if (earlyAlert.getCreatedBy() == null) {
    throw new ValidationException("CreeatedBy is missing.");
  }

  if (earlyAlert.getCampus() == null) {
    throw new ValidationException("Campus is missing.");
  }
}\end{lstlisting}
\end{minipage}

\scalebox{2}{$\Downarrow$}

\vspace{0.2cm}
\begin{minipage}{0.48\textwidth}
\begin{lstlisting}
public Map<String, Object> fillTemplateParameters(
  @NotNull final EarlyAlert earlyAlert) {

  EarlyAlertValidation.checkIsNull(earlyAlert);
  EarlyAlertValidation.checkCreatedBy(earlyAlert);
  EarlyAlertValidation.checkCampus(earlyAlert);
}\end{lstlisting}
\end{minipage}

\vspace{-0.2cm}
\caption{P2: Encapsulating business rules}
\vspace{-0.2cm}
\label{fig:p4}
\end{center}
\end{figure}

\vspace{0.2cm}
\noindent
\textbf{P2: Encapsulating business rules (14\% / 86\%).}
This pair is somehow similar to P1, although com rather different developers' perception. The original version had three \texttt{if}s statements testing conditions and raising an exception message if the conditions were met. The refactored version encapsulated each condition into a specific method, abstracting away the business logic.   %\gnote{The CDD approach reduced ICP by \xxx} 
%was created a class and specific methods for each IF. As a result of the questionnaires we obtained 19 votes in favor of the original code and 114 in favor of the refactored code. 
There were 19 votes in favor of the original version. One of these respondents mentioned that it is important to known which exception could be raise, while another respondent did not see any value in creating an specific validation class. 
%comments favored the original code because they considered it easier to be understood by Java beginners ("\textit{Both are readable, but for the size of the method and validations I don't see the benefit in creating a specific validation class.}"; "\textit{Because the way the if is being used in code 1 helps to understand the possible exceptions that may occur in the code, especially for java beginners.}")
On the other hand, the participants that favored the CDD version pointed out that the refactored version is simpler and intuitive. For instance, one respondent highlighted that \textit{the method signature is clear and it makes it easy to understand its behavior}", whereas another respondent said that  ``\textit{The second snippet, besides being leaner, may represent an opportunity for code reuse}". Indeed, several participants that the CDD version creates room for code reuse. Participants also 
%\textit{Isolation of behavior in methods that facilitates reuse.}";"\textit{Much simpler and intuitive}") the refactored approach was pointed out by several interviewees as a great possibility to reuse the code, they also 
highlighted the importance of the method signatures to be consistent with their operation, since it makes the code much simpler and intuitive.

\vspace{0.2cm}
\noindent
\textbf{P3: Concatenating logical expressions (7\% / 93\%).}
In this practice, we present a method with four \texttt{if} statements (three nested ones). In the CDD-version, the logical expressions of these four \texttt{if}s were grouped into a single \texttt{if} statement. %\gnote{The CDD approach reduced ICP by \xxx}
A total of 124 participants (93\%) voted for the CDD version as the most readable one. One CDD voter commented that this was the ``\emph{Classic example of cyclomatic complexity. I couldn't understand the example with several nested if's}''. Other respondents also commented that the excessive indentation of the nested \texttt{if}s hinders readability.
However, one participant that voted in the original version brought an interesting perspective: ``\emph{it is easier to debug the code when ifs are separated [...] It also eases the use of InteliJ features such as `evaluate expression'}''.

%As comments to be highlighted in favor of the refactored form with CDD ("\textit{The greater stringing of if's within if's hinders readability, so even though the validation of the second section is longer, it allows a better reading because it is unique}";" \textit{Classic example of cyclomatic complexity. I couldn't understand the example with several nested if's}";" \textit{The indentation of the first one is very bad.}";"\textit{ The indentation is readable, better written in my view.}") the vast majority of participants agree that multiple IF's hinder the traceability of the code and its understanding, another highlighted and relevant factor was the better indentation that the refactored way allowed to do.

\vspace{0.2cm}
\noindent
\textbf{P4: Extracting class (11\% / 89\%).}
In this practice, in the original code, we presented the implementation of a \texttt{Comparator} abstract class within the body method. On the other hand, in the CDD version, we introduced a new class, \texttt{MessageParamsComparator}, that implemented the \texttt{Comparator} behavior. The new class is accessed by using \texttt{MessageParamsComparator::compare} in the original method. 
%The practice called "Extracting class"described in P04 compares a code original method compares two objects, while the refactored code is a class where the comparator method was encapsulated in another class. 
The vast majority of the voters (89\%) favored the CDD version. 
%The result of the questionnaires was 14 votes in favor of the original code and 119 in favor of the refactored code. 
One CDD voter commented that 
%the issue of established habit and little knowledge of lambda functions was highlighted again, and method reference the highlights in favor of the refactored form with CDD (" 
``\emph{The separation into a new method, with a suitable name, makes the code more consistent and easier to understand}''. Another respondent commented that the CDD version ``\emph{Decreased cognitive complexity after extracting business logic}'', which is in sharp alignment with the CDD purpose. Still, one CDD voter also highlighted the potential for code reuse: ``\textit{Besides being more readable, it can be reusable}''.
%" \textit{More intuitive code and use of lambda and method reference}"; " \textit{Besides being more readable it can be reusable}") are related to the separation of responsibilities, the issue of self-explanatory method names, decreased cognitive complexity with the separation of responsibilities and the possibility of responsibilities.
Regarding the comments in favor of the original code, one respondent mentioned that ``\emph{It's the way I'm most used to do. But I consider it a `less elegant' way}''. Other respondent thought that ``\emph{the first option is clearer for those who don't know lambda functions}''.

\vspace{0.2cm}
\noindent

\textbf{P5: Listing all imports (34\% / 66\%).}
Here we compared a code snippet in which the imports are implicit (when we used (\texttt{*}) wildcard to hidden classes of the same package) and another (the CDD version) where the imports are explicit (when we listed all imports). %\gnote{The CDD approach reduced ICP by \xxx}  %As a result of the questionnaires, a balanced result was obtained of 45 votes in favor of the original code and 88 in favor of the refactored code. 
This question had some interesting comments. 
Most of voters favoring the original code believed that the implicit imports bring better readability because it uses fewer lines of code and, then, it makes it easier to identify groups of imports. In this regard, respondent mentioned that ``\textit{The shorter list is clearly more readable, although it might be importing classes that will not be used.}''. On the other hand, developers that advocate in favor of the CDD design argue that it is important to understand what is being imported, as one respondent said: ``\emph{Despite making the files more extensive, it is preferable to import classes individually, since it eases their location and understanding. Also, it does not import unnecessary things to the code}''. Some developers also mentioned that the explicit approach eases code maintenance, since it reduces the chances of importing the wrong class.

%" \textit{Easier to identify the imports, even if it is not explicit which files are being used}") most believed that the implicit imports bring better readability because it is written in fewer lines of code and easier to identify the imports and with respect to the highlights of the refactored version with CDD I highlight (" \textit{I like the explicit import because it makes clear everything I need; 

%"Despite making the files more extensive, it is preferable to perform the imports individually to facilitate their location and understanding, and not import unnecessary things to the code}"; " \textit{explicit control of what is imported, helps in code maintenance}") the comments that give greater control pro understanding of the programmer which classes are really being used this helps in code maintenance and implementation because it is not uncommon to have problems importing the wrong class.

\vspace{0.2cm}
\noindent
\textbf{P6: Encapsulating business rules (34\% / 66\%).}
This is the same kind of practice reported in P2. The original code had two conditionals to check if an array is not empty and, if not, inserts several elements at once in the array, using the \texttt{addAll} method. The CDD approach encapsulated the two \texttt{if} statements into two methods. %\gnote{The CDD approach reduced ICP by \xxx} 
%(\texttt{queriesTemplate.putIfEmpty} and \texttt{headersTemplate.putIfEmpty}). 
Similar to the results of P2, in this practice, the majority of the respondents (106 of them) also preferred the CDD version.
The reasons for (and against) CDD were also similar, as the ones provided in P2. For instance, those that favored the CDD approach commented that ``\emph{The reduction (or even encapsulation) of \texttt{if}s improves a lot the code readability}'' and ``\emph{The [newly introduced] method's name is self-explanatory}''. One final interesting observation was regarding the negation in the condition (i.e., \texttt{if (!requestTemplate.queries().isEmpty())}), as one participant highlighted: ``\emph{The negation used in the original condition increases cognitive complexity and hinders the code's understanding}''.

%The results of the questionnaires were 27 votes in favor of the original code and 106 in favor of the refactored code. 
%Regarding the comments in favor of the original code, it is emphasized that by habit the original form is preferable and also that in this case IF used with moderation can be better than a method that hides the responsibility ("\textit{It's the way I'm most used to doing it. But I consider it a "less elegant" way}"; "\textit{IFs, when used sparingly, can make the code more readable}"). Regarding the highlights of the refactored version with CDD I highlight ("\textit{I think the function names are very explanatory}"; "\textit{Using negation before the sentence I think is better only for simpler conditionals. And, again, in case there was already a method that does what is expected, the readability is much better.}"; "\textit{The method name is self-explanatory.}"; "The reduction (or even delegation) of ifs helps the readability of the code a lot. "The name of the method is very self-explanatory and makes it much easier to understand its function. Another highlighted factor is the negation of the function used in the original code, which increases cognitive complexity and makes understanding more difficult.

\begin{figure}[h]
\begin{center}
\begin{minipage}{0.48\textwidth}
\begin{lstlisting}
public Request request() {
  if (!this.resolved) {
    throw new IllegalStateException("template has not been resolved.");
  }
  return Request.create(method, url(), headers(), body, this);
}\end{lstlisting}
\end{minipage}

\scalebox{2}{$\Downarrow$}

\vspace{0.2cm}
\begin{minipage}{0.48\textwidth}
\begin{lstlisting}
public Request request() {
  Asserts.booleanStateMustBeTrue(this.resolved, "template has not been resolved.");
  return Request.create(method, url(), headers(), body, this);
}\end{lstlisting}
\end{minipage}

\vspace{-0.2cm}
\caption{P7: Asserting conditions}
\vspace{-0.2cm}
\label{fig:p7}
\end{center}
\end{figure}

\vspace{0.2cm}
\noindent
\textbf{P7: Asserting conditions (45\% / 55\%).}
This practice is somewhat similar to P2 and P6, in the sense that the \texttt{if} statement is encapsulated, but in here the CDD-approach used a native Java \texttt{Assert} method to perform the comparison and throw the exception (if needed). %\gnote{The CDD approach reduced ICP by \xxx} 
Developers were undecided, with a slight  advantage for the code refactored with CDD.
%The practice called "Asserting conditions" described in P07 compares a code that uses a comparison to demonstrate an exception and the refactored solution uses an Assert method (native JAVA), to perform the comparison and throw the exception. 
%The results of the questionnaires showed a slight advantage for the code refactored with CDD, the result was 60 votes in favor of the original code and 73 in favor of the refactored code. 
Regarding the comments in favor of the original code, one developer mentioned that the ``\emph{\texttt{Assert} methods are more conventional to testing code}''. Another one said that ``\emph{The if is simple, I don't see the need to extract it into a separate method}". 
%Considering that assert methods are widely used in unit tests, they didn't consider that this practice brought good readability to the example. 
On the other hand, those that favored CDD commented, once again, on the reuse opportunity (``\emph{Code reuse is guaranteed}". Another respondent complemented that ``\emph{The assertion makes explicit the expected condition (and what will happen if it isn't met).}".
%; "\textit{Good use of Asserts.}") that the refactored code guarantees reusability, besides for validation where it will only be executed if the condition is not met the Assert Method validation is much more adequate.

\begin{figure}[h]
\begin{center}
\begin{minipage}{0.48\textwidth}
\begin{lstlisting}
public String method() {
  return (method != null) ? method.name() : null;
}\end{lstlisting}
\end{minipage}

\scalebox{2}{$\Downarrow$}

\vspace{0.2cm}
\begin{minipage}{0.48\textwidth}
\begin{lstlisting}
public String method() {	
  return Optional.ofNullable(method)
                 .map(HttpMethod::name)
                 .orElse(null);
}\end{lstlisting}
\end{minipage}

\vspace{-0.2cm}
\caption{P8: Functional checking style}
\vspace{-0.2cm}
\label{fig:p8}
\end{center}
\end{figure}

\vspace{0.2cm}
\noindent
\textbf{P8: Functional checking style (71\% / 29\%).}
Here we present a validation using a ternary operator and the CDD-version that followed a functional style using \texttt{Optional}. %\gnote{The CDD approach reduced ICP by \xxx} 
This was the only case in which the original code was vastly preferred. 
Those that voted for CDD mentioned that (``\textit{Reflects more the OO paradigm}'')
Regarding the original code, developers commented that the \texttt{map} function was brought too much complexity (``\textit{The ternary is an easy resource and there was no complex logic to justify a \texttt{.map}}''). One respondent also mentioned that the ternary operator was of better use ``\emph{than chaining \texttt{if}s statements}''.

%The practice called "Functional style checking if var is null" described in P08 compares a validation solved by a ternary comparison and the refactored code uses a lambda function from JAVA. The results of the questionnaires showed a considerable advantage for the original code: 95 votes were in favor of the original code and 38 in favor of the refactored code. 
%Regarding the comments favorable to the original code, the following are highlighted ("\textit{The ternary is an easy resource to recognize and there was no logic complex enough to justify a .map}"; "\textit{simpler for those who don't understand optional}"; "\textit{In this case, because it is a simple verification, the ternary conditional is more readable than the use of map + lambda"). Better than stringing ifs (or ternaries)}; "\textit{the method already makes it explicit and there's no need to interpret a conditional}"). Analyzing the comments both in favor of and against factoring, one can see that in simpler situations the use of ternary makes the code more readable, despite recognizing the better efficiency of lambda functions.

\begin{figure}[h]
\begin{center}
\begin{minipage}{0.48\textwidth}
\begin{lstlisting}
public String url() {
  StringBuilder url = new StringBuilder(this.path());
  if (!this.queries.isEmpty()) {
    url.append(this.queryLine());
  }
  if (fragment != null) {
    url.append(fragment);
  }

  return url.toString();
}\end{lstlisting}
\end{minipage}

\scalebox{2}{$\Downarrow$}

\vspace{0.2cm}
\begin{minipage}{0.48\textwidth}
\begin{lstlisting}
public String url() {
  return PathBuilder.withAllQueryParameters(this);
}
...
public class PathBuilder {
    public static String withAllQueryParameters(RequestTemplate requestTemplate) {
    StringBuilder url = new StringBuilder(requestTemplate.path());
    if (!requestTemplate.getQueriesTemplate().isEmpty()) {
      url.append(requestTemplate.queryLine());
    }
    if (requestTemplate.getFragment().isPresent()) {
      url.append(requestTemplate.getFragment().get());
    }

    return url.toString();
  }
}\end{lstlisting}
\end{minipage}

\vspace{-0.2cm}
\caption{P9: Extracting class}
\vspace{-0.2cm}
\label{fig:p9}
\end{center}
\end{figure}

\vspace{0.2cm}
\noindent
\textbf{P9: Extracting class (33\% / 67\%).}
This is the second occurrence of this kind of pair. In here, the original code used a \texttt{StringBuilder} and \texttt{if} statements to concatenate URL elements. The CDD version moved the \texttt{StringBuilder} and the \texttt{if} statements into a newly introduced class (called \texttt{PathBuilder}), and the URL is built using \texttt{PathBuilder.withAllQueryParameters(this)} (Figure~\ref{fig:p9}). %\gnote{The CDD approach reduced ICP by \xxx} 
The results of the questionnaires showed a slight advantage for the code refactored with CDD version (89/\totalSurveyParticipants developers voted for this solution).
One of the participants that favored the CDD version said that this version ``\textit{encapsulates the logic, and makes it easier to read. If necessary I enter the method to understand what it does.}''.
Interestingly, one developer that favored the original code had the opposite idea: ``\textit{Unless there is \texttt{PathBuilder} reuse, the code in \texttt{url()} will end up the same way inside the \texttt{withAllQueryParameters} method. The developer, in a sense, will need to go inside this method to see what is being done. It just increased one more layer to get to the code, since the \texttt{url()} method wouldn't have any other logic or flow.}"

%The practice called "Extracting class" described in P09 compares a code that forms a URL in the original version uses the StringBuilder which builds the URL using comparisons. 
%The results of the questionnaires showed a slight advantage for the code refactored with CDD, the result was 44 votes in favor of the original code and 89 in favor of the refactored code. 
%Regarding comments in favor of the original code, the following is highlighted ("\textit{Unless there is PathBuilder reuse, the code in url() will end up the same way inside the withAllQueryParameters method. The developer, in a sense, will need to go inside this method to see what is being done. It just increased one more layer to get to the code, since the url() method wouldn't have any other logic or flow. In Favor of refactoring I highlight ("Better code division and less complexity regarding the use of conditionals}"; "\textit{Encapsulates the logic, and makes it easier to read. If necessary I enter the method to understand what it does.}"; "Separation of responsibilities helps readability.") the division of cognitive complexity that is one of the pillars of the CDD made the code easier to understand.

\vspace{0.2cm}
\noindent
\textbf{P10: Encapsulating for loops (8\% / 92\%).}
In this final example, the original code concatenated added elements into a list using two \texttt{for} loops and one \texttt{if} statement, whereas the CDD version abstracted the two \texttt{for}s by using one native method of the \texttt{List} interface. Those in favor of the CDD approach mentioned that ``\emph{Even not knowing the whole API, just by the name of the method it is easy to understand its purpose}'' and \textit{Concatenating lists makes more sense and it is cleaner to use a lambda and an \texttt{addAll} than doing a \texttt{for} or \texttt{foreach}. The idea is the same, we use less line, and, we don't have to maintain this piece of code}''.

In general, when considering the 10 pairs of code snippets chosen for this study it is possible to observe the influence of a cognitive constraint on reducing the presence of ICPs in all refactored classes. It is important to note that CDD focuses on improving code units, i.e., the classes (in object-oriented languages) are the main structures to apply the CDD principles. For this reason, we do not count in such code snippets the number of ICPs before and after refactorings because we would need to consider the whole class. However, this reduction can be perceived as if the CDD soft-forced the developers to restructure the classes to meet a satisfactory understanding threshold.

%The results of the questionnaires showed a slight advantage for the code refactored with CDD, the result was 10 votes in favor of the original code and 123 in favor of the refactored code. 
%With regard to favorable comments, the habit of using FOR and greater clarity of what is being done were highlighted. On the other hand, the developers in favor of refactoring by CDD highlighted ("\textit{Even not knowing the whole API, just by the name of the method it is easy to identify its purpose}"; "\textit{By using the existing methods the code is cleaner and easier to understand its purpose}"; "\textit{Concatenating lists makes more sense and it is cleaner to use a lambda and an \texttt{addAll} than doing a for, \texttt{foreach}. The idea is the same, we have less line, and, hardly will have a maintenance in this piece of code}") after the refactoring even not knowing what the method does in full became more readable due to the method name being self explanatory and improving the understanding and readability.

\subsection*{RQ2. \RQtwo}\label{sec:rq2}

In here we present our findings regarding our second research question. Table~\ref{tab:model} describes the results after applying Posnett model. In this table there are some $\times$~symbols that indicate that we were unable to calculate the metric. This happened due to two reasons. First, the pairs P4 and P9 moved part of the code to a newly introduced class.
As we mentioned in Section~\ref{sec:model}, the tool we used to calculate Posnett model does not process an entire Java class, only method bodies. Since P4 and P9 contains class declarations, we were unable to run the tool on them.
We were also unable to run the tool in P5 (which compares two approaches for listing imports), because Java methods do not accept \texttt{imports} declarations.

\begin{table}[t!]
    \centering
    \caption{The results of Posnett model. We calculate the number of lines of code (LOC column) using the \texttt{wc} UNIX tool (considering blank lines). The symbol $\times$~indicates when it was not possible to calculate it.} %\gnote{atualizar o P9 pois o LOC da versão refatorada está sendo calculado incorretamente, so leva em consideração a assinatura da class} 
    \label{tab:model}
    \begin{tabular}{lcccc}
    \toprule
    Pairs & LOC Before & LOC After & Posnett Before & Posnett After  \\
    \midrule
    P1 & 9 & 5 & 0.0208 & 0.0174 \\
    P2 & 13 & 6 & 0.0013 & 0.0023 \\
    P3 & 16 & 11 & 0.1205 & 0.3368 \\
    P4 & 16 & 16 & 0.7029 & $\times$ \\
    P5 & 5 & 10 & $\times$ & $\times$ \\
    P6 & 12 & 7 & 0.0053 & 0.0092 \\
    P7 & 5 & 3 & 0.0192 & 0.0304 \\
    P8 & 2 & 2 & 0.0149 & 0.0276\\
    P9 & 10 & 10 & 0.0066 & $\times$ \\
    P10 & 9 & 5 & 0.0127 & 0.0253 \\
    \bottomrule
    \end{tabular}
\end{table}

This table shows a couple of interesting observations. First, we could perceived that for 9 out of the 10 \cddguided, the CDD version had fewer lines of code, when compared to the original version; on average, the CDD versions used 37\% less lines of code. The only exception is P5, which doubled the number of lines of code used. This happened because the CDD version adopted the \emph{explicit} approach (which lists all imports), while the original version used the \emph{implicit} approach (which hide some of the imports), using the asterisk (\texttt{*}) wildcard.

Moreover, regarding the results of the Posnett model, we noticed that CDD excelled in only one out of the seven pairs that we were able to calculate the metric; P1, in that case (the model performance was 0.0174, when compared to 0.0208 of the original version). For the all other pairs of code snippets, the original version had better performance then the \cddguided. For this metric, the lower the value, the better readability the code snippet has. 

However, when taking a closer look at this result, we could observe that the model performance did not varied much between the pairs of code snippets. For instance, for P1, P2, P6, P8, and P10, the performance variation was less than $\sim$0.01. On the other hand, P3 and P7 showed a larger performance variation. For P3, in particular, the \cddguided combined four \texttt{if} statements into a single \texttt{if} with several logical expressions. It is important to note that the Posnett model is, in part, based on the diversity of the code vocabulary, that is, the sum of unique operators and unique operands. Since the CDD-refactored version reduced the number of \texttt{if} statements, the operands were not reused anymore among the \texttt{if} statements. This might have worsened its performance due to the higher number of unique operators and operands.

\section{Discussions}

Cognitive-Driven Development (CDD) is a coding technique that aims to reduce code complexity by always aiming to reduce the developer's cognitive load. These strategies are not fixed and can be adapted to each development team. The research in question evaluates refactorings based on CDD. However, although most of the practices are favorable to refactoring, we cannot affirm that CDD by itself is responsible for improving readability. However, it favors using other refactoring practices and good practices of Object-Oriented Programming.

The P2 is evidenced in Figure~\ref{fig:p4}, where the name of the method, being self-explanatory, proved to be a very efficient practice, being mentioned by approximately 30 of the 124 participants in favor of the CDD approach. They highlighted in their comments that the method's name made all the difference in refactoring, making the language very natural and easy to understand.
On the other hand, practice P2 (Figure~\ref{fig:p4}) fostered reuse and was a point identified by the research participants.  Similarly, the use of stable classes such as \texttt{List} and \texttt{Assert} favor the use of CDD as they naturally decrease the number of ICPs. We also observed that CDD-inspired code snippets are smaller than the original ones.

As a final observation, we noticed that CDD makes the code more horizontally aligned. It happens because most of the refactorings in the code result in new methods or classes that, when reaching a limit, continue to be modified to fit the human mind, i.e., a satisfactory understanding degree.

\section{Related Work}

We group our works in terms of 1) empirical studies that aims to assess code readability (Section~\ref{sec:readability}) and 2) early work on CDD (Section~\ref{sec:cdd-related}).

\subsection{Assessing Code Readability}\label{sec:readability}

%\gnote{temos que organizar essa seção. agrupar em termos de contextos dos trabalhos. descrever como esses trabalhos se diferem dos nossos?}

%We consider readability as a fundamental concept for code maintenance and evolution. Within maintenance work, programmers spend about 70\% of their time trying to understand the source code, which could decrease if the code was more readable~\cite{CK:Scalabrino2017}. Readability refers to the ease with which a person can read and understand a piece of code~\cite{CK:BuseWeimer}.

Several researches have been working on the topic of code readability. While some proposed coding standards and conventions to assess  code readable, others related readability to the cognitive and complexity metrics. We next discuss some of the closest works to ours.
%Although there are no previous works discussing the improvement or worsening of readability after code refactoring using the CDD, there are studies that discuss the effect of other coding patterns on software projects.

\vspace{0.2cm}
\noindent
\textbf{Empirical studies on code readability.}
Works evaluating the readability of code snippets have already been done by researchers.
Gerosa et al.~\cite{Santos:ICPC:2018} conducted a survey with software developers evaluating coding convention patterns and showed that most of these patterns positively influence the readability perceived by developers. A survey was conducted with 55 students of the software engineering course of the Computer Science course and another smaller group formed by 7 professional programmers of a large Brazilian software company. Buse and Weimer's~\cite{Buse:TSE:2009} practices and the Scalabrino's model~\cite{CK:Scalabrino2017} were evaluated. At the end of the analysis it was identified that 8 out of 11 coding practices affected the readability perceived by the research participants. In our study, 10 pairs of code snippets were chosen from a previous study involving refactorings based on CDD.
%Other work aims to quantify relationships between code attributes and developers' perceived readability, with the goal of creating models to measure readability.
Sivaprakasam et al.~\cite{Sivaprakasam:2012} provided a tool to support their proposed method, which takes java methods themselves as input and returns refactored, readable source code by inserting blank lines after each block of valid code. Experimental results of the research showed that the automatic insertion of blank lines left the code with relevant lightness and a better understanding of the snippet due to the standardized organization and spacing of the parts with less cognitive load leaving the code more pleasant to read.

\vspace{0.2cm}
\noindent
\textbf{Descriptive model for code readability}
In the study by Buse and Weimer \cite{Buse:TSE:2009}, readability was defined as the human understanding of the ease of understanding a text and that the readability of a program is related to its maintainability. The hypothesis of the study is that programmers have some intuitive notion to point out program features and characteristics that will be good indicators for readability. With this, a descriptive model of software readability based on programmers' opinions and notions of software quality was presented. To construct the model Buse and Weimer they conducted a study with 120 students of different levels of coding experience, asking participants to provide subjective rating scores of reading code snippets. A Survey was conducted where each participant was given the same set of snippets. Participants could select a number close to five for ``most readable'' snippets or close to one for ``least readable'' snippets, with a score of three indicating neutrality. The result of this study was a set of code snippets accompanied by 12,000 evaluations on readability. 

Posnett et al. \cite{Posnett:MSR:2011} extended the Buse and Weimer's model \cite{Buse:TSE:2009} providing a simple and intuitive readability theory based on code size and entropy. They also point out some details that may have negatively influenced the Buse and Weimer model. They performed an extraction of code snippets longer than 200 lines and compared the two models addressed. As a result, it was possible to show that the proposed model outperforms the Buse and Weimer model as a readability classifier in small code snippets.

In 2018, Mannan et al. \cite{Mannan:2018} used Posnett's readability metric to evaluate readability in large open-source projects. However, the results found a very low correlation between source code bad smells and readability. Considering the results and that Posnett's model was initially evaluated with small projects, Mannan et al. concluded that there are deficiencies in current readability models and therefore, there is a need to identify better metrics to evaluate readability.

In 2018 Fakhoury et al.\cite{Fakhoury:ICPC:2018}, explored the effect of poor source code lexicon and readability on cognitive load as measured by a state-of-the-art minimally invasive functional imaging technique called functional near-infrared spectroscopy (fNIRS). The research results evidenced a significant increase in participants' cognitive load when anti-linguistic patterns and structural inconsistencies were introduced to the source code; for passages considered more readable, there was a decrease in cognitive load overload.

\vspace{0.2cm}
\noindent
\textbf{Programming language features and code elements}
In a work by Mendonça et al.\cite{Mendonca:JSERD:2020}, the impact of lambdas functions on JAVA programmers' understanding is evaluated. There is a common understanding that code refactoring with lambdas functions, in addition to other potential benefits, simplifies code and improves program understanding. A survey was carried out with 158 pairs of code snippets extracted from GitHub. As a result of the work after comparing with the Buse and Weimer model \cite{Buse:TSE:2009} and Posnett.\cite{Posnett:MSR:2011} a contradictory result was found, both models suggested that refactoring by lambda functions does not improve the understandability of the source code, however in the qualitative result (survey) indicated that the introduction of lambda expressions in legacy code improves the understanding of the code in particular cases.

%\gnote{falar do trabalho da Fernanda e do Castor} \cite{Oliveira:ICSME:2020}

\subsection{Early Work on CDD}\label{sec:cdd-related}

Souza and Pinto~\cite{CDD:ICSME:2020} described the concepts that support the CDD. Continuous expansion is part of the complex nature of software. However, the understandability cannot follow in the same proportion. The lack of a clear relationship between software complexity and program comprehension contributes to the software not evolving healthy. As a consequence, software developers spend a considerable part of their time on program understanding~\cite{minelli2015}. Developers must know when to restructure the code and possibly improve the separation of responsibilities. According to the authors, when we do not have a well-defined rule for intrinsic complexity for source code, it will be increasingly common to find classes that contribute to a cognitive overload for developers.

Pinto at al.\cite{CDD:ENASE:2021} (detailed in \ref{sec:refactoring}) reached the conclusion that refactorings using conventional practices guided by a complexity constraint were better evaluated when they were compared with the refactoring clusters (all classes created or modified) without such rule. 

Pereira at al. \cite{CDD:SBES:2021} provided a tool called ``Cognitive Load Analyzer'' to support the CDD, a plugin for IntelliJ IDEA and Java language. The intrinsic complexity of the code is calculated through static analysis during programming, and the tool observes the limit of complexity. When the complexity limit is reached for some code unit, a notification is displayed to suggest possible refactorings.

In recent research, Pinto and Souza \cite{CDD:ICEIS:2022} evaluated the effects of adopting a complexity constraint in the early stages of software development. Three projects adopted by some companies for hiring new software engineers were selected to be developed by 44 experienced developers, divided into Non-CDD and CDD groups. Both groups were aware of the importance of quality metrics and the need to produce high-quality code for other developers to understand. The CDD group received different training that included practices guided by a cognitive complexity limit, including suggestions for elements to set a constraint. The result suggested that CDD can guide the developers to achieve lower dispersion for the quality metric measures (CK metrics).

The concept of Cognitive-Driven Development opens the door for extensive experimental research to measure the effectiveness of this strategy regarding the measurement of complexity from source code. Although most of the works discussed here involve the program understanding, qualitative studies involving readability criteria have not yet been carried out.

\section{Limitations}

As any empirical work, our also have limitations and threats to validity. 

First, we spent several weeks cleaning and filtering the data provided by Pinto and colleagues~\cite{CDD:ENASE:2021}. Despite our best efforts to find representative \cddguided (see details at Section~\ref{sec:refactoring}), we may not have selected a diverse set of code snippets. To mitigate this threat, after grouping the code snippets into categories, we sought to have at most two code snippets per category. 

Some participants also pointed out that some of our examples were very simple and, thus, not adequate. It is worth noting that such simplicity was intentional to make the survey feasible and less tiring. If we provide more complex examples that require more cognitive effort, we could potentially discourage participants from answering the survey. Still regarding the code snippets, manually selecting and formatting the code snippets may have, in some way, influenced the opinions of the participants. For example, by showing the code snippet as an image instead of a text, we may have limited the way our participants interacted with the examples (for instance, there were not able to copy the code to their IDEs), negatively influencing the opinions of some research participants.

Another limitation is regarding the way we present the code snippets. In the daily development routine, developers rarely have to spend time reading short methods without further navigation and navigation. Therefore, our experiment hardly assembles the real world development routine. We believe that experiments such as this one could help researchers and tool builders create better models that could, in turn, be used to guide developers in writing software of better quality.

Finally, our data is based on the responses provided by \totalSurveyParticipants professional software developers. Although these developers belong to a group of great interest to our researcher, that is, professional developers with extensive software development experience, these developers pertain to a relatively restricted group of professional programmers with experience in Java. Therefore, it is not possible to generalize the conclusions obtained to general groups of programmers.

\section{Conclusions}

In this paper we evaluate the extent to which CDD improves code readability. To achieve this goal, we conducted a two-phase study. We conducted a survey with professional software developers invited to vote (and justify their reasoning) for the most readable pair of code snippets (out of a set of 10 pairs); one of the pairs was coded using CDD practices. We received 133 responses.
In the second phase, we applied the state-of-the-art readability model to the 10 pairs of CDD-guided refactorings. We observed some with conflicting results.

The results allowed us to answer the research question RQ1, the developers perceived that seven (out of 10) CDD-guided refactorings were more readable than their counterparts; for two other CDD-guided refactorings the developers were undecided, while for only one of the CDD-guided refactorings did the developers prefer the original code snippet. 
Regarding research question RQ2, we note that only two CDD-driven refactorings show better readability, as evaluated by state-of-the-art readability models~\cite{Posnett:MSR:2011}. 

We conclude that by following CDD design practices, coding units are expected to be smaller and thus easier to maintain and evolve and thus easier to understand. It is important to stress that CDD only indicates a decrease in intrinsic Class complexity, so other refactoring methods are combinatorial and favor readability from CDD.Our results provide initial evidence that CDD may be an interesting approach to software design.

As perspectives for future work, we can perform the analysis by other readability methods and serve as a basis for their evolution, because our results demonstrate sometimes conflicting positions about some situations on the same piece of code and sometimes a near unanimity in favor of a piece of code. This information can be valuable to encourage other studies in the area.

%\section{Acknowledgments}

\bibliographystyle{ACM-Reference-Format}
\bibliography{references}


\begin{thebibliography}{24}

%%% the Web address.

\ifx \showCODEN    \undefined \def \showCODEN     #1{\unskip}     \fi
\ifx \showDOI      \undefined \def \showDOI       #1{#1}\fi
\ifx \showISBNx    \undefined \def \showISBNx     #1{\unskip}     \fi
\ifx \showISBNxiii \undefined \def \showISBNxiii  #1{\unskip}     \fi
\ifx \showISSN     \undefined \def \showISSN      #1{\unskip}     \fi
\ifx \showLCCN     \undefined \def \showLCCN      #1{\unskip}     \fi
\ifx \shownote     \undefined \def \shownote      #1{#1}          \fi
\ifx \showarticletitle \undefined \def \showarticletitle #1{#1}   \fi
\ifx \showURL      \undefined \def \showURL       {\relax}        \fi
% invisible to TeX
\providecommand\bibfield[2]{#2}
\providecommand\bibinfo[2]{#2}
\providecommand\natexlab[1]{#1}
\providecommand\showeprint[2][]{arXiv:#2}

\bibitem[Buse and Weimer(2009)]%
        {Buse:TSE:2009}
\bibfield{author}{\bibinfo{person}{Raymond~PL Buse} {and}
  \bibinfo{person}{Westley~R Weimer}.} \bibinfo{year}{2009}\natexlab{}.
\newblock \showarticletitle{Learning a metric for code readability}.
\newblock \bibinfo{journal}{\emph{IEEE Transactions on software engineering}}
  \bibinfo{volume}{36}, \bibinfo{number}{4} (\bibinfo{year}{2009}),
  \bibinfo{pages}{546--558}.
\newblock


\bibitem[C.~M.~Chung and Yang(2010)]%
        {Chung:2010}
\bibfield{author}{\bibinfo{person}{W.~R.~Edwards C.~M.~Chung} {and}
  \bibinfo{person}{M.~G. Yang}.} \bibinfo{year}{2010}\natexlab{}.
\newblock \showarticletitle{Static and Dynamic Data Flow Metrics}. In
  \bibinfo{booktitle}{\emph{Policy and Information}},
  Vol.~\bibinfo{volume}{13}. \bibinfo{publisher}{{IEEE}},
  \bibinfo{pages}{1--6}.
\newblock


\bibitem[Chidamber and Kemerer(1994)]%
        {CK:TSE}
\bibfield{author}{\bibinfo{person}{Shyam~R Chidamber} {and}
  \bibinfo{person}{Chris~F Kemerer}.} \bibinfo{year}{1994}\natexlab{}.
\newblock \showarticletitle{A metrics suite for object oriented design}.
\newblock \bibinfo{journal}{\emph{IEEE Transactions on software engineering}}
  \bibinfo{volume}{20}, \bibinfo{number}{6} (\bibinfo{year}{1994}),
  \bibinfo{pages}{476--493}.
\newblock


\bibitem[de~Souza and Pinto(2020)]%
        {CDD:ICSME:2020}
\bibfield{author}{\bibinfo{person}{Alberto Luiz Oliveira~Tavares de Souza}
  {and} \bibinfo{person}{Victor Hugo Santiago~Costa Pinto}.}
  \bibinfo{year}{2020}\natexlab{}.
\newblock \showarticletitle{Toward a Definition of Cognitive-Driven
  Development}. In \bibinfo{booktitle}{\emph{{IEEE} International Conference on
  Software Maintenance and Evolution, {ICSME} 2020, Adelaide, Australia,
  September 28 - October 2, 2020}}. \bibinfo{publisher}{{IEEE}},
  \bibinfo{pages}{776--778}.
\newblock
\urldef\tempurl%
\url{https://doi.org/10.1109/ICSME46990.2020.00087}
\showDOI{\tempurl}


\bibitem[dos Santos and Gerosa(2018)]%
        {Santos:ICPC:2018}
\bibfield{author}{\bibinfo{person}{Rodrigo~Magalh{\~{a}}es dos Santos} {and}
  \bibinfo{person}{Marco~Aur{\'{e}}lio Gerosa}.}
  \bibinfo{year}{2018}\natexlab{}.
\newblock \showarticletitle{Impacts of coding practices on readability}. In
  \bibinfo{booktitle}{\emph{Proceedings of the 26th Conference on Program
  Comprehension, {ICPC} 2018, Gothenburg, Sweden, May 27-28, 2018}},
  \bibfield{editor}{\bibinfo{person}{Foutse Khomh},
  \bibinfo{person}{Chanchal~K. Roy}, {and} \bibinfo{person}{Janet Siegmund}}
  (Eds.). \bibinfo{publisher}{{ACM}}, \bibinfo{pages}{277--285}.
\newblock


\bibitem[Fakhoury(2018)]%
        {Fakhoury:ICPC:2018}
\bibfield{author}{\bibinfo{person}{Ma~Y. Arnaoudova V. \& Adesope-O. Fakhoury,
  S.}} \bibinfo{year}{2018}\natexlab{}.
\newblock \showarticletitle{The effect of poor source code lexicon and
  readability on developers’ cognitive load}. In \bibinfo{booktitle}{\emph{In
  Proceedings of the 26th Conference on Program Comprehension - ICPC ’18 New
  York, New York, USA: ACM Press.}} \bibinfo{publisher}{{ACM}},
  \bibinfo{pages}{286--296}.
\newblock


\bibitem[Halstead(1977)]%
        {halstead1977elements}
\bibfield{author}{\bibinfo{person}{Maurice~H Halstead}.}
  \bibinfo{year}{1977}\natexlab{}.
\newblock \bibinfo{booktitle}{\emph{Elements of Software Science (Operating and
  programming systems series)}}.
\newblock \bibinfo{publisher}{Elsevier Science Inc.}
\newblock


\bibitem[Hannebauer et~al\mbox{.}(2012)]%
        {Sivaprakasam:2012}
\bibfield{author}{\bibinfo{person}{Christoph Hannebauer}, \bibinfo{person}{Marc
  Hesenius}, {and} \bibinfo{person}{Volker Gruhn}.}
  \bibinfo{year}{2012}\natexlab{}.
\newblock \showarticletitle{An accurate model of software code readability}.
\newblock \bibinfo{journal}{\emph{International Journal of Engineering Research
  and Technology. ESRSA Publications}} \bibinfo{volume}{1}, \bibinfo{number}{6}
  (\bibinfo{year}{2012}).
\newblock


\bibitem[Hannebauer et~al\mbox{.}(2018)]%
        {hannebauer2018does}
\bibfield{author}{\bibinfo{person}{Christoph Hannebauer}, \bibinfo{person}{Marc
  Hesenius}, {and} \bibinfo{person}{Volker Gruhn}.}
  \bibinfo{year}{2018}\natexlab{}.
\newblock \showarticletitle{Does syntax highlighting help programming novices?}
\newblock \bibinfo{journal}{\emph{Empirical Software Engineering}}
  \bibinfo{volume}{23}, \bibinfo{number}{5} (\bibinfo{year}{2018}),
  \bibinfo{pages}{2795--2828}.
\newblock


\bibitem[Kitchenham and Pfleeger(2008)]%
        {kitchenham2008personal}
\bibfield{author}{\bibinfo{person}{Barbara~A Kitchenham} {and}
  \bibinfo{person}{Shari~L Pfleeger}.} \bibinfo{year}{2008}\natexlab{}.
\newblock \showarticletitle{Personal opinion surveys}.
\newblock In \bibinfo{booktitle}{\emph{Guide to advanced empirical software
  engineering}}. \bibinfo{publisher}{Springer}, \bibinfo{pages}{63--92}.
\newblock


\bibitem[Mendonça et~al\mbox{.}(2020)]%
        {Mendonca:JSERD:2020}
\bibfield{author}{\bibinfo{person}{Walter Lucas~Monteiro Mendonça},
  \bibinfo{person}{José Fortes}, \bibinfo{person}{Francisco~Vitor Lopes},
  \bibinfo{person}{Diego Marcílio}, \bibinfo{person}{Rodrigo~Bonifácio de
  Almeida}, \bibinfo{person}{Edna~Dias Canedo}, \bibinfo{person}{Fernanda
  Lima}, {and} \bibinfo{person}{João Saraiva}.}
  \bibinfo{year}{2020}\natexlab{}.
\newblock \showarticletitle{Understanding the Impact of Introducing Lambda
  Expressions in Java Programs}.
\newblock \bibinfo{journal}{\emph{Journal of Software Engineering Research and
  Development}}  \bibinfo{volume}{8} (\bibinfo{date}{Oct.}
  \bibinfo{year}{2020}), \bibinfo{pages}{7:1 – 7:22}.
\newblock
\urldef\tempurl%
\url{https://doi.org/10.5753/jserd.2020.744}
\showDOI{\tempurl}


\bibitem[Mi(2018)]%
        {MiQ:2018}
\bibfield{author}{\bibinfo{person}{Keung J. Xiao Y. Mensah S.\& Gao~Y. Mi, Q.}}
  \bibinfo{year}{2018}\natexlab{}.
\newblock \showarticletitle{Improving code readability classification using
  convolutional neural networks}. In \bibinfo{booktitle}{\emph{Information and
  Software Technology,}}. \bibinfo{pages}{60--71}.
\newblock
\urldef\tempurl%
\url{https://linkinghub.elsevier.com/retrieve/pii/S0950584918301496}
\showURL{%
\tempurl}


\bibitem[Miller(1956)]%
        {miller1956magical}
\bibfield{author}{\bibinfo{person}{George~A Miller}.}
  \bibinfo{year}{1956}\natexlab{}.
\newblock \showarticletitle{The magical number seven, plus or minus two: Some
  limits on our capacity for processing information.}
\newblock \bibinfo{journal}{\emph{Psychological review}} \bibinfo{volume}{63},
  \bibinfo{number}{2} (\bibinfo{year}{1956}), \bibinfo{pages}{81}.
\newblock


\bibitem[Minelli et~al\mbox{.}(2015)]%
        {minelli2015}
\bibfield{author}{\bibinfo{person}{Roberto Minelli}, \bibinfo{person}{Andrea
  Mocci}, {and} \bibinfo{person}{Michele Lanza}.}
  \bibinfo{year}{2015}\natexlab{}.
\newblock \showarticletitle{I know what you did last summer-an investigation of
  how developers spend their time}. In \bibinfo{booktitle}{\emph{2015 IEEE 23rd
  International Conference on Program Comprehension}}. IEEE,
  \bibinfo{pages}{25--35}.
\newblock


\bibitem[Oliveira et~al\mbox{.}(2020)]%
        {Oliveira:ICSME:2020}
\bibfield{author}{\bibinfo{person}{Delano Oliveira}, \bibinfo{person}{Reydne
  Bruno}, \bibinfo{person}{Fernanda Madeiral}, {and} \bibinfo{person}{Fernando
  Castor}.} \bibinfo{year}{2020}\natexlab{}.
\newblock \showarticletitle{Evaluating Code Readability and Legibility: An
  Examination of Human-centric Studies}. In \bibinfo{booktitle}{\emph{{IEEE}
  International Conference on Software Maintenance and Evolution, {ICSME} 2020,
  Adelaide, Australia, September 28 - October 2, 2020}}.
  \bibinfo{publisher}{{IEEE}}, \bibinfo{pages}{348--359}.
\newblock
\urldef\tempurl%
\url{https://doi.org/10.1109/ICSME46990.2020.00041}
\showDOI{\tempurl}


\bibitem[Opdyke(1990)]%
        {opdyke1990}
\bibfield{author}{\bibinfo{person}{William~F Opdyke}.}
  \bibinfo{year}{1990}\natexlab{}.
\newblock \showarticletitle{Refactoring: An aid in designing application
  frameworks and evolving object-oriented systems}. In
  \bibinfo{booktitle}{\emph{Proc. SOOPPA'90: Symposium on Object-Oriented
  Programming Emphasizing Practical Applications}}.
\newblock


\bibitem[Pereira et~al\mbox{.}(2021)]%
        {CDD:SBES:2021}
\bibfield{author}{\bibinfo{person}{Jherson Haryson~A. Pereira},
  \bibinfo{person}{Alberto Luiz Oliveira~Tavares de Souza}, {and}
  \bibinfo{person}{Victor Hugo Santiago~C. Pinto}.}
  \bibinfo{year}{2021}\natexlab{}.
\newblock \showarticletitle{Cognitive Load Analyzer: {A} Support Tool for
  Cognitive-Driven Development}. In \bibinfo{booktitle}{\emph{{SBES} '21: 35th
  Brazilian Symposium on Software Engineering, Joinville, Santa Catarina,
  Brazil, 27 September 2021 - 1 October 2021}},
  \bibfield{editor}{\bibinfo{person}{Cristiano~D. Vasconcellos},
  \bibinfo{person}{Karina~Girardi Roggia}, \bibinfo{person}{Vanessa Collere},
  {and} \bibinfo{person}{Paulo Bousfield}} (Eds.). \bibinfo{publisher}{{ACM}},
  \bibinfo{pages}{468--473}.
\newblock
\urldef\tempurl%
\url{https://doi.org/10.1145/3474624.3476011}
\showDOI{\tempurl}


\bibitem[Pinto et~al\mbox{.}(2021)]%
        {CDD:ENASE:2021}
\bibfield{author}{\bibinfo{person}{Victor Hugo Santiago~C. Pinto},
  \bibinfo{person}{Alberto Luiz Oliveira~Tavares de Souza},
  \bibinfo{person}{Yuri Matheus~Barboza de Oliveira}, {and}
  \bibinfo{person}{Danilo~Monteiro Ribeiro}.} \bibinfo{year}{2021}\natexlab{}.
\newblock \showarticletitle{Cognitive-Driven Development: Preliminary Results
  on Software Refactorings}. In \bibinfo{booktitle}{\emph{Proceedings of the
  16th International Conference on Evaluation of Novel Approaches to Software
  Engineering, {ENASE} 2021, Online Streaming, April 26-27, 2021}},
  \bibfield{editor}{\bibinfo{person}{Raian Ali}, \bibinfo{person}{Hermann
  Kaindl}, {and} \bibinfo{person}{Leszek~A. Maciaszek}} (Eds.).
  \bibinfo{publisher}{{SCITEPRESS}}, \bibinfo{pages}{92--102}.
\newblock
\urldef\tempurl%
\url{https://doi.org/10.5220/0010408100920102}
\showDOI{\tempurl}


\bibitem[Pinto and Tavares(2022)]%
        {CDD:ICEIS:2022}
\bibfield{author}{\bibinfo{person}{Victor Hugo Santiago~C. Pinto} {and}
  \bibinfo{person}{Alberto Luiz~Oliveira Tavares}.}
  \bibinfo{year}{2022}\natexlab{}.
\newblock \showarticletitle{Effects of Cognitive-driven Development in the
  Early Stages of the Software Development Life Cycle}. In
  \bibinfo{booktitle}{\emph{Proceedings of the 24th International Conference on
  Enterprise Information Systems - Volume 2, Online Streaming, April 25-27,
  2022}}. \bibinfo{publisher}{{SCITEPRESS}}, \bibinfo{pages}{40--51}.
\newblock


\bibitem[Posnett et~al\mbox{.}(2011)]%
        {Posnett:MSR:2011}
\bibfield{author}{\bibinfo{person}{Daryl Posnett}, \bibinfo{person}{Abram
  Hindle}, {and} \bibinfo{person}{Premkumar~T. Devanbu}.}
  \bibinfo{year}{2011}\natexlab{}.
\newblock \showarticletitle{A simpler model of software readability}. In
  \bibinfo{booktitle}{\emph{Proceedings of the 8th International Working
  Conference on Mining Software Repositories, {MSR} 2011 (Co-located with
  ICSE), Waikiki, Honolulu, HI, USA, May 21-28, 2011, Proceedings}},
  \bibfield{editor}{\bibinfo{person}{Arie van Deursen}, \bibinfo{person}{Tao
  Xie}, {and} \bibinfo{person}{Thomas Zimmermann}} (Eds.).
  \bibinfo{publisher}{{ACM}}, \bibinfo{pages}{73--82}.
\newblock


\bibitem[S.~Scalabrino and Oliveto(2010)]%
        {CK:Scalabrino2017}
\bibfield{author}{\bibinfo{person}{C.~Vendome M. Linares-Vasquez D.~Poshyvanyk
  S.~Scalabrino, G.~Bavota} {and} \bibinfo{person}{R. Oliveto}.}
  \bibinfo{year}{2010}\natexlab{}.
\newblock \showarticletitle{Automatically assessing code understandability: How
  far are we?}
\newblock \bibinfo{journal}{\emph{in 32nd IEEE/ACM International Conference on
  Automated Software Engineering (ASE)}} (\bibinfo{year}{2010}),
  \bibinfo{pages}{417--427}.
\newblock


\bibitem[Sweller(1988)]%
        {sweller1988}
\bibfield{author}{\bibinfo{person}{John Sweller}.}
  \bibinfo{year}{1988}\natexlab{}.
\newblock \showarticletitle{Cognitive load during problem solving: Effects on
  learning}.
\newblock \bibinfo{journal}{\emph{Cognitive science}} \bibinfo{volume}{12},
  \bibinfo{number}{2} (\bibinfo{year}{1988}), \bibinfo{pages}{257--285}.
\newblock


\bibitem[Sweller(2010)]%
        {sweller2010}
\bibfield{author}{\bibinfo{person}{John Sweller}.}
  \bibinfo{year}{2010}\natexlab{}.
\newblock \showarticletitle{Cognitive load theory: Recent theoretical
  advances.}
\newblock  (\bibinfo{year}{2010}).
\newblock


\bibitem[U.~A.~Mannan and Sarma(2011)]%
        {Mannan:2018}
\bibfield{author}{\bibinfo{person}{I.~Ahmed U.~A.~Mannan} {and}
  \bibinfo{person}{A. Sarma}.} \bibinfo{year}{2011}\natexlab{}.
\newblock \showarticletitle{Towards understanding code readability and its
  impact on design quality}. In \bibinfo{booktitle}{\emph{in Proceedings of the
  4th ACM SIGSOFT International Workshop on NLP for Software Engineering}}.
  \bibinfo{publisher}{{IEEE}}, \bibinfo{pages}{18--21}.
\newblock
\urldef\tempurl%
\url{https://doi.org/10.1145/3283812.3283820}
\showURL{%
\tempurl}


\end{thebibliography}

\end{document}